\documentclass
[twocolumn,superscriptaddress,showpacs,prb,amsfonts,amssymb,final,balancelastpage]{revtex4}%
\usepackage{amsmath}
\usepackage{amssymb}
\usepackage{graphicx}
\usepackage{times}
\usepackage{amsfonts}%
\providecommand{\U}[1]{\protect\rule{.1in}{.1in}}

\begin{document}
\title{Intrinsic spin-Hall accumulation in honeycomb lattices: Band structure effect}
\author{Ming-Hao Liu}
\homepage{http://homepage.ntu.edu.tw/~d92222010/}
\email{d92222010@ntu.edu.tw}
\affiliation{Department of Physics, National Taiwan University, Taipei 10617, Taiwan}
\author{Gustav Bihlmayer}
\affiliation{Institut f\"{u}r Festk\"{o}rperforschung, Forschungszentrum J\"{u}lich,
D-52425 J\"{u}lich, Germany}
\author{Stefan Bl\"{u}gel}
\affiliation{Institut f\"{u}r Festk\"{o}rperforschung, Forschungszentrum J\"{u}lich,
D-52425 J\"{u}lich, Germany}
\author{Ching-Ray Chang}
\affiliation{Department of Physics, National Taiwan University, Taipei 10617, Taiwan}
\received{24 July 2007}

\published{5 September 2007}

\eid{sss}

\begin{abstract}
Local spin and charge densities on a two-dimensional honeycomb lattice are
calculated by the Landauer-Keldysh formalism (LKF). Through the empirical
tight-binding method, we show how the realistic band structure can be brought
into the LKF. Taking the Bi(111) surface, on which strong surface states and
Rashba spin-orbit coupling are present [Phys. Rev. Lett. \textbf{93}, 046403
(2004)], as a numeric example, we show typical intrinsic spin-Hall
accumulation (ISHA) patterns thereon. The Fermi-energy-dependence of the spin
and charge transport in two-terminal nanostructure samples is subsequently
analyzed. By changing $E_{F}$, we show that the ISHA pattern is nearly
isotropic (free-electron-like) only when $E_{F}$ is close to the band bottom,
and is sensitive/insensitive to $E_{F}$ for the low/high bias regime with such
$E_{F}$. With $E_{F}$ far from the band bottom, band structure effects thus
enter the ISHA patterns and the transport direction becomes significant.

\end{abstract}
\maketitle

In electron systems, the extrinsic spin-Hall effect (SHE), theoretically
proposed long time ago,\cite{DPspinHall,Hirsch} has been experimentally proven
optically in semiconductor bulk structures\cite{Awschalom science}\ and
two-dimensional electron systems (2DESs),\cite{Awschalom 2DEG} and even
electrically in diffusive metallic conductors.\cite{ElectricSpinHallExp}
Recent achievement on observing the SHE in 2DESs at room
temperatures\cite{RTspinHall} has further enlightened the possibility to
manipulate spins via the SHE based on such a simple mechanism: transverse spin
separation by\ passing through longitudinal electric currents.

As for the intrinsic SHE, theoretically proposed much later than the extrinsic
one,\cite{Sinova} experimental evidence for its existence has been achieved
only in two-dimensional hole systems\cite{Wunderlich} but not in 2DESs. In
particular, the local spin scanning in real-space for intrinsic SHE systems is
difficult to carry out due to its limited size, and hence the required extra
high resolution. Contrary to the extrinsic type, the intrinsic spin-Hall
accumulation (SHA) pattern shows not only the out-of-plane component of spin
accumulating antisymmetrically at the two lateral sides near the sample edges,
but also oscillations due to the wave function modulation. Moreover, the SHA
pattern may vary with, e.g., bias strength, spin-orbit coupling (SOC)
strength, sample size and shape.\cite{Nikolic,Shenger}

\begin{figure}
[b]
\begin{center}
\includegraphics[
height=1.8464in,
width=3.2353in
]%
{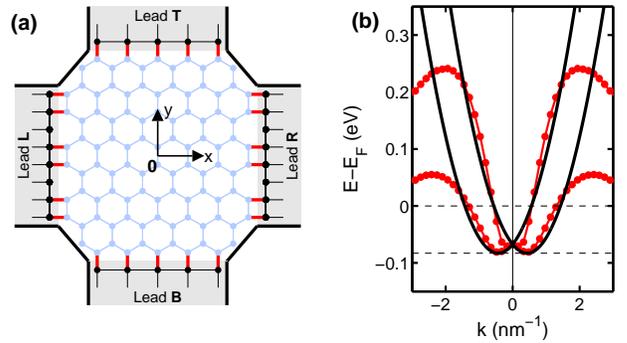}%
\caption{(Color online) (a) Schematic of the four-terminal setup with the
conducting sample made of honeycomb lattice structure. (b) Band structures
calculated by first principle (red dots) and by empirical TBM (black lines).}%
\label{fig1}%
\end{center}
\end{figure}

In this paper we further investigate the crystal-structure-dependence, and
hence the band structure effect, of the intrinsic SHA (ISHA) pattern in 2DESs.
The honeycomb lattice structure is particularly suitable for such
investigation due to its nontrivial band structure and interesting geometry.
In addition, recent confirmation of the strong Rashba SOC\cite{Rashba} on Bi
surfaces,\cite{Hofmann review} in particular the (111) case,\cite{Koroteev
2004} in which the projected bilayer structure exactly forms the honeycomb
lattice, provides a good numerical example to demonstrate these effects. Based
on the Landauer-Keldysh formalism (LKF),\cite{Nikolic} the local spin
densities (LSD) on four-terminal nanostructure samples made of the honeycomb
lattice [see Fig. \ref{fig1}(a)], are calculated. We will show that the ISHA
pattern (i) exhibits isotropic/anisotropic spin transport behaviors when the
Fermi level is near/away from the band bottom, (ii) shows dramatic difference
between the left-right (zigzag) and the bottom-top (armchair) transport modes
when the Fermi level lies in the band gap, and (iii) is extremely sensitive to
the Fermi level in the low-bias regime.%

To apply the LKF, the first step is to construct the real space
tight-binding-like Hamiltonian, which builds the underlying band structure and
therefore is decisive for the transport properties. For the honeycomb lattice
and considering both kinetic and Rashba hoppings up to the nearest neighbor
only, the model Hamiltonian can be written as\cite{Kane-Mele}%
\begin{equation}
\mathcal{H}=\sum_{i}\varepsilon_{i}c_{i}^{\dag}c_{i}+\sum_{\left\langle
ij\right\rangle }c_{j}^{\dag}\left[  -t_{0}-it_{R}\left(  \vec{\sigma}%
\times\mathbf{d}_{ji}\right)  _{z}\right]  c_{i}, \label{H}%
\end{equation}
with $c_{i}^{\dag}$ ($c_{i}$) being the creation (annihilation) operator of
the electron on site $i$. Here the first term is the on-site energy with
parameter $\varepsilon_{i},$ which may also describe the local potential and
the disorder. In our potential-free case here, $\varepsilon_{i}$ simply
corresponds to the band energy offset in the language of tight-binding model
(TBM). The second term in Eq. (\ref{H}) contains the kinetic and Rashba
hoppings with strengths $t_{0}$ and $t_{R}$, respectively. In the Rashba
hopping term, $\mathbf{d}_{ji}$ is the unit vector pointing from site $i$ to
site $j$, and $\vec{\sigma}\equiv\left(  \sigma^{x},\sigma^{y},\sigma
^{z}\right)  $ is the Pauli matrix vector.

To obtain realistic values for the above parameters $\varepsilon_{i},$
$t_{0},$ and $t_{R},$ we next perform the empirical TBM\cite{GrossoBook} of
the Slater-Koster type,\cite{Slater-Koster} based on Eq. (\ref{H}).
Considering the two triangular sublattices forming the honeycomb lattice and
taking only single orbital $p_{z}$ on each site into account, it can be shown
that the Hamiltonian matrix equivalent to Eq. (\ref{H}) reads%
\begin{equation}
\mathbb{H}=\left(
\begin{array}
[c]{cc}%
\mathbb{H}_{11} & \mathbb{H}_{12}\\
\mathbb{H}_{12}^{\dag} & \mathbb{H}_{11}%
\end{array}
\right)  , \label{Hmatrix}%
\end{equation}
with the diagonal element given by $\mathbb{H}_{11}=E_{p}\mathbb{I}_{2},$
where $E_{p}$ and $\mathbb{I}_{2}$ are the $p$-orbital energy and the
$2\times2$ identity matrix, respectively, and the off-diagonal element given
by%
\begin{equation}
\mathbb{H}_{12}=\left(
\begin{array}
[c]{cc}%
U\left(  1+2F\right)  & -it_{R}\left(  1-F-\sqrt{3}G\right) \\
-it_{R}\left(  1-F+\sqrt{3}G\right)  & U\left(  1+2F\right)
\end{array}
\right)  , \label{H12}%
\end{equation}
where $U\equiv l_{z}^{2}V_{pp\sigma}+\left(  1-l_{z}^{2}\right)  V_{pp\pi}$ is
the two-center interaction integral involving $p_{z}$ atomic orbitals, and the
compact functions are given by $F\equiv\exp(-i\sqrt{3}k_{y}a/2)\cos\left(
k_{x}a/2\right)  $ and $G\equiv\exp(-i\sqrt{3}k_{y}a/2)\sin\left(
k_{x}a/2\right)  .$ Note that for an ideal (flat) two-dimensional honeycomb
lattice such as graphene, only the $\pi$ bands contribute to $U$ due to the
vanishing direction cosine $l_{z}=0$. However, later we will choose the
Bi(111) bilayer structure as a numerical example, in which both $\sigma$ and
$\pi$ bands contribute due to the nonvanishing $l_{z}$. Still, here in both
cases only the composite parameter $U$ is to be extracted, instead of the
individual $V_{pp\sigma}$ and $V_{pp\pi}$. Except the Rashba hopping strength
$t_{R}$ which we denoted consistently in both Eq. (\ref{H}) and the empirical
TBM Eqs. (\ref{Hmatrix}) and (\ref{H12}), the other two parameters are related
by $-t_{0}=U$ and $\varepsilon_{i}=E_{p}.$ Diagonalizing the $4\times4$ matrix
of Eq. (\ref{Hmatrix}) gives the two pairs of the four energy dispersion curves.

To extract reasonable parameters for Eq.~(\ref{H}), we consider the Bi(111)
bilayer structure, for its strong surface states,\cite{Hofmann review} making
the electron transport 2DES-like, and its strong Rashba SOC,\cite{Koroteev
2004} making the ISHE thereon promising. Considering nearest-neighbor hopping
only (and hence neglecting the interbilayer hopping), the projected
two-dimensional lattice structure is exactly of honeycomb type. We therefore
fit the energy dispersion curves obtained from diagonalizing
Eq.\ (\ref{Hmatrix}) with the surface band structure from the first principle
calculation,\cite{Koroteev 2004} as shown in Fig.~\ref{fig1}(b). Both
directions in this plot are along $\bar{\Gamma}\bar{M}$. Due to the simple
model we have taken, the fitting gives good agreement only near the
$\bar{\Gamma}$ point. Correspondingly, the band parameters extracted from such
fitting are $t_{0}=1.6302%
\operatorname{eV}%
,$ $t_{R}=0.1853%
\operatorname{eV}%
,$ and $\varepsilon_{i}=4.8324%
\operatorname{eV}%
.$

The lead-sample setup is sketched in Fig. \ref{fig1}(a), where the sample made
of the honeycomb lattice is of nearly square shape with four terminals (left,
right, bottom, and top), each of which can be contacted by a semi-infinite
normal metal lead. Throughout the rest of the calculations, we will set the
$x$ and $y$ axes along the zigzag and armchair directions, respectively, and
fix the origin of $\left(  x,y\right)  =\left(  0,0\right)  $ at the center of
the sample, as indicated in Fig. \ref{fig1}(a).%
\begin{figure}
[ptb]
\begin{center}
\includegraphics[
height=2.2649in,
width=2.8279in
]%
{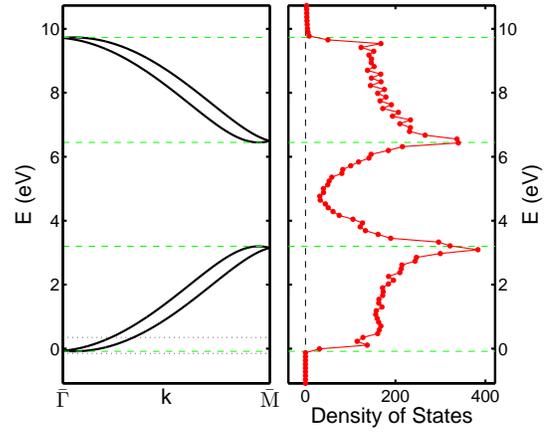}%
\caption{(Color online) Left panel: the band structure along $\bar{\Gamma}%
\bar{M}$ direction calculated by the empirical TBM. Right panel: the total
density of states of the sample with size about $4.1\times4.2$
$\operatorname{nm}^{2}$ calculated by the LKF. The dashed lines in both panels
are guided for the eyes and the dotted lines in the left panel indicate the
energy range shown in Fig. \ref{fig1}(b).}%
\label{fig2}%
\end{center}
\end{figure}

The LKF, namely, the nonequilibrium Keldysh Green's function
formalism\cite{Keldysh} applied on the Landauer multiterminal setups, has been
summarized in some detail in Ref. \onlinecite{Nikolic}. Its central spirit is
to solve the kinetic equation for the Keldysh Green's function matrix
$\mathbb{G}^{<}\left(  E\right)  $, written in the real-space representation
at energy $E$. Physical quantities such as local charge density (LCD), LSD, or
even the charge and spin current densities, can then be extracted from
$\mathbb{G}^{<}\left(  E\right)  $. In systems free of phase-breaking
interactions (such as electron-electron or electron-phonon interactions), the
self-energy modifying the carrier life time inside the sample is contributed
only from the contact leads through the nearest neighbor hopping with strength
set equal to $t_{0}$ [see the bold (red) connection lines in Fig.
\ref{fig1}(a)], and the Keldysh Green's function can be solved
exactly.\cite{DattaBook} Explicitly, the self-energy can be expressed as the
product of $t_{0}^{2}$ and the retarded surface Green's function of the
attached semi-infinite leads.\cite{Shenger,DattaBook} In our analysis, we will
concentrate on $z$-component of the LSD given by\cite{Nikolic}%
\begin{equation}
\left\langle S_{\mu}\right\rangle _{\mathbf{r}_{i}}=\frac{\hbar}{2}\int
_{E_{F}-eV_{0}/2}^{E_{F}+eV_{0}/2}\operatorname{Tr}\left[  \sigma^{\mu
}\mathbb{G}^{<}\left(  E;\mathbf{r}_{i},\mathbf{r}_{i}\right)  \right]  dE,
\label{<S>}%
\end{equation}
where $S_{\mu}\equiv\left(  \hbar/2\right)  \sigma^{\mu}$ is the $\mu$
component spin operator with $\mu=x,y,z$ and $\mathbf{r}_{i}$ is the position
vector of the $i$th lattice site. In the right-hand side, the $2\times2$
matrix $\mathbb{G}^{<}\left(  E;\mathbf{r}_{i},\mathbf{r}_{i}\right)  $ is the
$i$th diagonal submatrix element of the whole $\mathbb{G}^{<}\left(  E\right)
$, $E_{F}$ is the Fermi energy to be tuned in the later analyses, $e$ is the
electron charge, and $V_{0}$ is the applied potential difference between the
negatively and positively biased leads. The LCD, which will be shown to
modulate the LSD in the later analysis, is given by%
\begin{equation}
\left\langle eN_{e}\right\rangle _{\mathbf{r}_{i}}=e\int_{E_{F}-eV_{0}%
/2}^{E_{F}+eV_{0}/2}\operatorname{Tr}\mathbb{G}^{<}\left(  E;\mathbf{r}%
_{i},\mathbf{r}_{i}\right)  dE, \label{<eN>}%
\end{equation}
where $N_{e}$ is the electron number operator. Note that here the band bottom
$E_{b}$, which is set to the lowest energy of the full band structure
calculated from the previously introduced TBM, does not explicitly enter the
expressions (\ref{<S>}) and (\ref{<eN>}) but will play a decisive role in
dealing with the self-energy due to the lead coupling. It may become even more
crucial when calculating physical quantities to which equilibrium states also
contribute, such as the bond spin current density.\cite{Nikolic}

Before performing the LSD and LCD calculations, one last step is to report the
convincing correspondence between the band structure of the infinitely
extending honeycomb lattice, calculated by the empirical TBM, and the total
density of states (DOS) of the finite-size sample, calculated by the LKF. The
latter is given by summing the spectral function for each site $\rho
_{T}\left(  E\right)  =-i\sum_{i}\operatorname{Tr}\left\{  \operatorname{Im}%
\left[  \mathbb{G}^{R}\left(  \mathbf{r}_{i},\mathbf{r}_{i};E\right)  \right]
\right\}  /\pi,$ where $\mathbb{G}^{R}\left(  \mathbf{r}_{i},\mathbf{r}%
_{i};E\right)  $ is the spin-resolved $i$th diagonal submatrix element of the
retarded Green's function matrix. As shown in Fig. \ref{fig2}, the main
features, including the band top, band gap, and band bottom, are nicely
correspondent with each other. Note that the drop of the total DOS at the band
gap region cannot be perfectly step-function-like since the sample considered
in the LKF is not infinitely large.%

\begin{figure}
[ptb]
\begin{center}
\includegraphics[
height=4.574in,
width=2.8072in
]%
{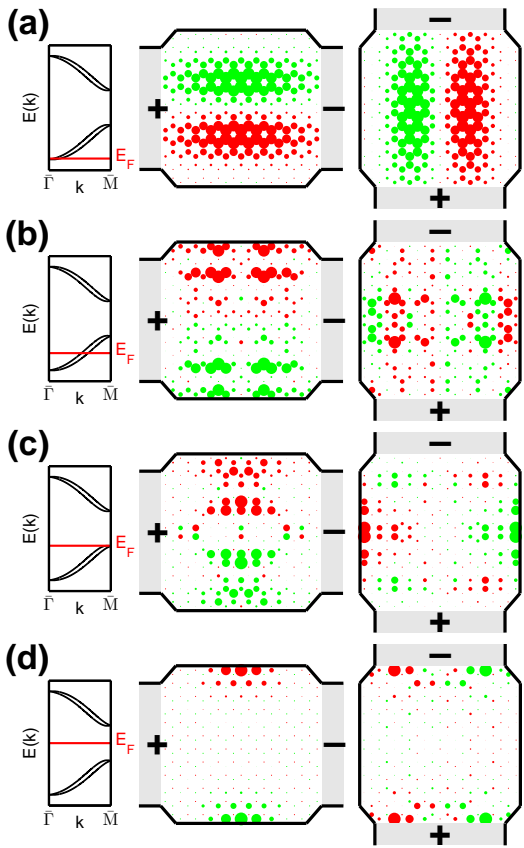}%
\caption{(Color online) Local spin density on a $2.3\times2.2$
$\operatorname{nm}^{2}$ sample with low bias with the Fermi level $E_{F}$ set
at (a) $0.08$ $\operatorname{eV}$ above the band bottom, (b) the middle of the
lower band, (c) the top edge of the lower band, and (d) the middle of the
energy gap. The red/green (dark/light) dots represent positive/negative
$\left\langle S_{z}\right\rangle $ with the dot size proportional to the
magnitude. The maximum values of $\left\langle S_{z}\right\rangle $ are of the
order of $10^{-6}\left(  \hbar/2\right)  $ in (a), and $10^{-4}\left(
\hbar/2\right)  $ for (b)--(d).}%
\label{fig3}%
\end{center}
\end{figure}

Having constructed the correspondence between the empirical TBM and the LKF,
and extracted reasonable parameters by fitting to the first principles
calculation for the Bi(111) surface, we are now ready to present the local
spin, and later also charge, densities on the conducting sample in the
honeycomb lattice. Two bias regimes will be distinguished: low and high,
standing for bias voltages of $eV_{0}=0.002$ and $0.2$ $%
\operatorname{eV}%
$, respectively, in the rest of the paper. Although there are totally four
terminals free to contact the electrodes, we will consider only two-lead cases
with head-to-tail orientation, either parallel to the $x$ or $y$ axes.

We first present the LSD for a sample of totally $248$ lattice sites (sample
area about $2.3\times2.2$ $%
\operatorname{nm}%
^{2}$) in the low bias regime. To examine the direction dependence of the spin
transport, we set $E_{F}$ at some representative positions. As shown in Fig.
\ref{fig3}, we gradually raise $E_{F}$ from the band bottom ($E_{F}=0$, i.e.,
about $0.083$ $%
\operatorname{eV}%
$ above the bottom of the band, consistent with the first principle
calculation) to the middle of the band gap (about $E_{F}=4.82$ $%
\operatorname{eV}%
$). In each panel of Fig. \ref{fig3}, the $\left\langle S_{z}\right\rangle $
distribution is antisymmetric about the bias axis, and therefore exhibits the
main feature of the intrinsic SHE. Furthermore, as can be clearly seen that,
only when $E_{F}$ is set at the band bottom [Fig. \ref{fig3}(a)], the pattern
becomes nearly isotropic, i.e., the spin transport does not show direction
dependence and thus behaves as free electrons. With the increase of $E_{F}$,
Figs. \ref{fig3}(b) and \ref{fig3}(c) show distinct LSD distributions for the
two different transport modes. Interestingly, when $E_{F}$ is set in the
middle of the gap [Fig. \ref{fig3}(d)], a dramatic difference between the two
transport modes emerges. In the left-to-right case spin accumulation occurs at
the lateral edges, while in the bottom-to-top case there is almost nothing at
the lateral edges but some spots induced near the leads. This can be
understood by observing that at this Fermi energy the transport inside the
sample is supported by the edge states, which are contributed mostly from the
zigzag edges. Thus in the bottom-to-top geometry the whole sample behaves
nearly insulating: neither charge nor spin can pass through the sample.%

\begin{figure}
[ptb]
\begin{center}
\includegraphics[
height=3.1004in,
width=3.1756in
]%
{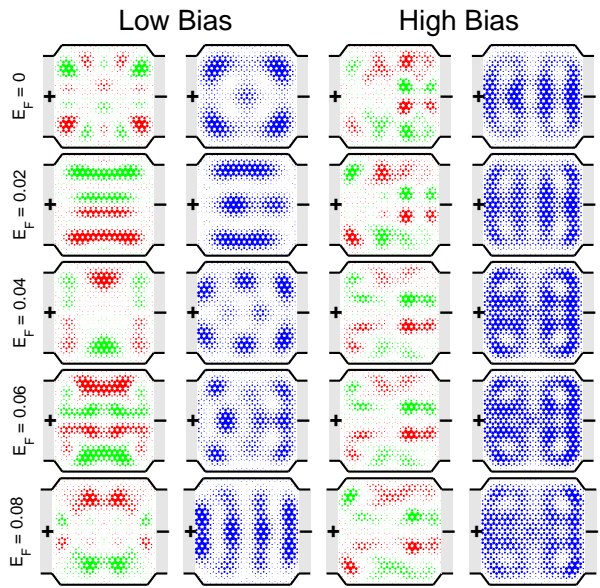}%
\caption{(Color online) LSD (the first and third columns) and LCD (the second
and fourth columns) on a $4.1\times4.2$ $\operatorname{nm}^{2}$ sample. Each
row corresponds to a gradually changing $E_{F}$ near the band bottom. The
left/right two columns are for low/high bias regimes. For LSD, the maximum
value of $\left\langle S_{z}\right\rangle $ in each panel in the low and high
bias regimes is of the order of $10^{-5}\left(  \hbar/2\right)  $ and
$10^{-3}\left(  \hbar/2\right)  $, respectively. For LCD, the maximum values
of $\left\langle N_{e}\right\rangle $ in each panel in the low and high bias
regimes are of the order of $10^{-4}$ and $10^{-3}$, respectively.}%
\label{fig4}%
\end{center}
\end{figure}

Next we take further looks at the spin and charge transport near $E_{F}=0$. We
consider a sample with 802 sites (about $4.1\times4.2$ $%
\operatorname{nm}%
^{2}$) with two leads in the left-to-right orientation, and finely raise the
Fermi level from $E_{F}=0$ $%
\operatorname{eV}%
$ to $E_{F}=0.08$ $%
\operatorname{eV}%
$. Both the low and high bias regimes will be analyzed. As shown in the first
column of Fig. \ref{fig4}, the ISHA pattern in the low-bias regime is
extremely sensitive to the Fermi level. Compared to the LCD in the second
column of Fig. \ref{fig4}, one can see that such sensitivity stems from the
wave function modulation. Whereas the nonequilibrium transport is contributed
from the states between $E_{F}-eV_{0}/2$ and $E_{F}+eV_{0}/2$, the low bias
regime with small voltages is mainly described by the Fermi energy state, and
the electron transport thus behaves quantum mechanically. It turns out that
the Fermi level, determining the length of the wave vector $k$, influences the
formation of the electron wave, and hence in turn the ISHA pattern. It is
interesting to note that a strong accumulation of electrons does not
necessarily lead to a prominent accumulation of spins. Conversely, where there
are no electrons, there must be no spins. Put in another way, the local spin
and charge density patterns must be, to some extent, consistent with each other.

For the high bias case, both the local spin and charge densities change
moderately with the increase of $E_{F}$ (the two right columns in Fig.
\ref{fig4}). Contrary to the low bias regime, the involved states
participating electron transport cover a much wider range of energy. Summation
of these states with different wave lengths eventually gives a waveless charge
distribution, as compared to the low-biased patterns. The electron transport
behavior is therefore far from the standard quantum-mechanical description. In
this case the nonequilibrium spin accumulation is no longer affected by the
wave function modulation, and the ISHA pattern is robust against the change of
$E_{F}$.

Before closing, it is worthy to remark here that in the pioneering formulation
of Ref. \onlinecite{Nikolic}, and also the recent application of Ref.
\onlinecite{Shenger} on the triangular lattices, the crystal structure
information is lost since the Fermi energy is chosen close to the band bottom:
$E_{F}=E_{b}+0.2t_{0}$. Consequently, the spin accumulation property remains
free-electron-like and exhibits rotational invariance (except for a
coexistence of the Dresselhaus term, giving rise to anisotropic dispersion).
It is only when the Fermi level is far from the band bottom, where the
corresponding wave vectors are short, that the band structure (or the crystal
structure) effect emerges.

In conclusion, taking the honeycomb lattice as a particular case, we have
pointed out the crucial role $E_{F}$ plays in the ISHE due to band structure
effects. Recent observation of the strong surface state and giant Rashba SOC
on Bi(111) surface\cite{Koroteev 2004} has attracted much attention and is
especially suitable as a numeric example in our investigation. The possibility
of observing the quantum SHE on the Bi(111) surface and its multilayer thin
film\cite{MurakamiBi} has made Bi(111) even more promising. Here we have
reported another positive viewpoint of its potential of observing the ISHE
thereon, provided that $E_{F}$ and the transport direction are important.
Moreover, we have combined the LKF with the first principle band calculation
through the empirical TBM, allowing one to extract realistic band parameters
for the LSD calculation.

One of us (M.H.L.) appreciates the hospitality in Forschungszentrum
J\"{u}lich, where most of this work was performed. Financial support of the
Republic of China National Science Council Grant No. 95-2112-M-002-044-MY3 is
gratefully acknowledged.

\end{document}